\documentclass[twocolumn,superscriptaddress,floats,prl,nofootinbib]{revtex4-1}
\usepackage{amssymb,amsmath,amssymb,amsfonts,amsthm,stmaryrd,mathrsfs,physics,bbm,mfirstuc}
\usepackage{graphicx}
\usepackage[T1]{fontenc}
\usepackage{enumerate} % advanced enumerate environment
\usepackage{color} % for colored text
\usepackage{graphpap} % for numbered coordinate grid
\usepackage[dvipsnames]{xcolor}

\usepackage{hyperref} % for HyperTeX cross-referencing

\newcommand{\heff}{{H_{\rm eff}}}
\newcommand{\F}{P}

\newcommand{\BNU}{School of Physics and Astronomy, \mbox{Key Laboratory of Multiscale Spin Physics (Ministry of Education)}, Beijing Normal University, Beijing 100875, China}

\newcommand{\mysection}[1]{\emph{\makefirstuc{#1}}---}
\newcommand{\myappendix}[1]{\mysection{\numappendix{#1}}}

\newcounter{mylabelno}

\newcommand{\mylabel}[1]{\refstepcounter{mylabelno}\label{#1}}

\newcounter{appendix}
\renewcommand{\theappendix}{Appendix \Alph{appendix}}
\newcommand{\numappendix}[1]{\refstepcounter{appendix}\textit{\theappendix}{}}

\begin{document}

\title{Covariant dynamics from static spherically symmetric geometries}

\author{Cong Zhang}
\email{cong.zhang@bnu.edu.cn}
\author{Zhoujian Cao}
\email{zjcao@bnu.edu.cn}
\affiliation{\BNU}

\begin{abstract}
This work reveals a fundamental link between general covariance and Birkhoff's theorem.
We extend Birkhoff's theorem from general relativity to a broad class of generally covariant gravity theories formulated in the Hamiltonian framework. Conversely, we show that each one-parameter family of static, spherically symmetric spacetimes determines a class of covariant theories, each of which has that family of spacetimes as its entire vacuum solution space.
Our systematic and model-independent framework applies to a wide range of spacetimes, including observationally inferred, quantum-gravity-inspired, and regular black holes. It provides a universal tool for probing their dynamical origins and enables the reconstruction of the underlying covariant theories from observational data, including gravitational-wave and black-hole-shadow measurements.
%observational tests of covariant gravitational dynamics through signals such as gravitational waves and black hole shadows.
\end{abstract}

%\keywords{Black hole, loop quantum gravity, singularity resolution }

\maketitle

%\tableofcontents

\mysection{Introduction}Recent advances in observational techniques \cite{LIGOScientific:2016aoc,LIGOScientific:2017bnn,EventHorizonTelescope:2019dse,EventHorizonTelescope:2022wkp} have made it increasingly feasible to probe static, spherically symmetric spacetimes that deviate from the Schwarzschild solution, the unique spherically symmetric vacuum solution of general relativity (GR), as guaranteed by Birkhoff's theorem. Against this backdrop, the inverse problem of Birkhoff's theorem arises naturally: how can one construct a self-consistent, covariant field-theoretic origin for a novel static, spherically symmetric spacetime beyond GR, such that the underlying theory admits the spacetime as its unique vacuum solution?

In addition to observational relevance, exploring  metrics that deviate from GR is theoretically motivated by addressing fundamental issues like spacetime singularities.
One common approach to generating such metrics involves ad hoc modifications of known black hole solutions \cite{bardeen1968non,Bronnikov:2000vy,Hayward:2005gi,Ansoldi:2008jw,Canate:2022zst,Lan:2023cvz}. The resulting geometries, known as regular black holes, are typically realized as solutions of GR coupled to exotic matter fields \cite{Ayon-Beato:1998hmi,Ayon-Beato:2000mjt,Bronnikov:2005gm,Bronnikov:2021uta,Chew:2022enh,Li:2024rbw}. However, this strategy is widely viewed as unsatisfactory, not only because it requires exotic matter, but also because the underlying theories generally continue to admit singular solutions \cite{Huang:2025uhv,Rao:2025rop}. Moreover, the absence of a dynamical framework that naturally predicts regular black holes has hindered systematic studies of their formation and stability, a topic that has recently attracted growing interest \cite{Bueno:2024eig,Harada:2025cwd,Vertogradov:2025snh,Bueno:2024zsx}.

Another line of research beyond GR seeks to incorporate quantum corrections to classical black hole solutions, aiming to resolve singularities through fundamentally modified gravitational dynamics. One strategy within this framework involves gluing a static, spherically symmetric exterior metric to a singularity-free interior model describing dust collapse in a quantum gravity (QG) framework \cite{Bambi:2013caa,Lewandowski:2022zce,Bonanno:2023rzk}. In such constructions, the exterior is typically determined by artificially introduced junction conditions, while the absence of a consistent dynamical description for the full spacetime remains a key limitation.  A  more systematic strategy that could overcome the limitations of these junction-based models is to directly quantize the spherically symmetric sector of GR \cite{Modesto:2008im,Gambini:2008dy,Corichi:2015xia,Ashtekar:2018lag,Husain:2021ojz,Gambini:2022hxr,Giesel:2023tsj,Ashtekar:2023cod,Belfaqih:2024vfk}. However, despite significant progress, these approaches have not yet converge to a unique, self-consistent picture of the underlying quantum-corrected dynamics. 

To systematically address the inverse problem of Birkhoff's theorem, a general and model-independent framework that ensures self-consistent dynamics is required.
The formalism developed in \cite{Zhang:2024khj,Zhang:2024ney} fulfills this need, as it derives from first principles a general form of gravity theories that restore general covariance within the Hamiltonian formulation, thereby providing an ideal theoretical foundation for the present study.

\mysection{Theoretical background}We start with the basic ingredients of the Hamiltonian framework for spherically symmetric GR on a 4-dimensional manifold $\mathcal M_2\times \mathbb S^2$. Here $\mathbb S^2$ denotes the 2-sphere,  and $\mathcal M_2$ is an arbitrary 2-dimensional Lorentzian  manifold of topology $\mathbb R\times\Sigma\ni (t,x)$, with $\Sigma$ being some 1-dimensional manifold. 
Due to the spherical symmetry, the theory is reduced to a dilaton gravity on $\mathcal M_2$ \cite{Grumiller:2002nm}. As shown in, e.g., \cite{Gambini:2013hna,Zhang:2021xoa}, the phase space of the dilaton gravity contains the canonical pairs $(K_2,E^2)$ for the 2-dimensional gravity and $(K_1,E^1)$ for the dilaton. From the 4-dimensional perspective, $E^I$ defines the spatial metric on the spatial manifold $\Sigma\times\mathbb S^2$, while $K_I$ gives its extrinsic curvature. 
The Poisson brackets of the canonical pairs read $\{K_1(x),E^1(y)\}=2\delta(x,y)$ and $\{K_2(x),E^2(y)\}=\delta(x,y)$, where geometrized units with $G=1=c$ are applied.

In classical GR, the 3+1 decomposition splits Einstein's equations into the Hamiltonian constraint, the diffeomorphism constraint, and the evolution equations. Since the evolution equations are equivalent to the Hamilton's equations generated by a linear combination of the constraints, the essence of the dynamics lies in these two constraints. In addition, the two constraints form a closed algebra under the Poisson bracket, constituting a first-class constraint system that guarantees the consistency between the constraint equations and Hamilton's equations.

we now extend all these classical structures to construct effective models that, for instance, could incorporate QG effects.  To go beyond Einstein's GR,  we assume that the Hamiltonian constraint is modified to an effective one, $\heff$, whose explicit form is yet to be determined, whereas the diffeomorphism constraint $H_x$ is taken to retain its classical expression,  $H_x=E^2\partial_xK_2-K_1\partial_xE^1/2$, which generates spatial diffeomorphisms. Consequently, the Poisson brackets involving $H_x$ are fully determined, and we only need to ensure the closure of the constraint algebra by considering the bracket $\{\heff[N], \heff[M]\}$ with $G[f]\equiv \int G f\dd x$. Since $\heff$ is modified, this bracket is expected to be deformed. The deformation is required to ensure that the constraints remain first-class and the constraint algebra mimics the structure of the hypersurface-deformation algebra, which facilitates the restoration of covariance.
Accordingly, we assume  $\{\heff[N], \heff[M]\} = H_x\left[\mu E^1/(E^2)^2(N\partial_x M - M\partial_x N)\right]$, where $\mu = 1$ in classical GR but, in general, may differ from unity and depend on phase-space variables.
 
 As a spacetime theory, it is necessary to introduce the metric as a spacetime tensor depending on the phase-space variables. To  guarantee that the deformed constraint algebra continues to admit the same geometric interpretation as in the classical theory, the metric is defined to be
\begin{equation}\label{eq:metric}
\dd s^2=-N^2\dd t^2+\frac{(E^2)^2}{\mu E^1}\left(\dd x+N^x\dd t\right)^2+E^1\dd\Omega^2,
\end{equation}
where $N$ and $N^x$ are known as the lapse function and the shift vector, and $\dd\Omega^2$ denotes the standard metric on the unit sphere $\mathbb S^2$.

To preserve the general covariance of the metric \eqref{eq:metric}, the effective Hamiltonian constraint $\heff$ must be restricted so that the gauge transformations it generates reproduce the diffeomorphism transformations of the metric \eqref{eq:metric}. This requirement imposes nontrivial conditions on $\heff$. In \cite{Zhang:2024khj,Zhang:2024ney}, these conditions were systematically analyzed, and a general form of $\heff$ consistent with general covariance was derived, 
\begin{equation}\label{eq:HeffOnMeff}
\heff=-2E^2\Big[\partial_{s_1}M_{\rm eff}+\frac{\partial_{s_2}M_{\rm eff}}{2}s_3+\frac{\partial_{s_4}M_{\rm eff}}{s_4}s_5+\mathcal R\Big],
\end{equation}
where $s_a$ for $a=1,2,\cdots,5$ are scalars defined by
\begin{equation*}
s_1=E^1,\ s_2=K_2,\ s_3=\frac{K_1}{E^2},\ s_4=\frac{\partial_xs_1}{E^2},s_5=\frac{\partial_xs_4}{E^2},
\end{equation*} 
$\mathcal R$ is an arbitrary function of $s_1$ and $M_{\rm eff}$, and the mass function $M_{\rm eff}$ depending on $s_1, s_2, s_4$ is a solution to the following so-called covariance equation:
\begin{subequations}\label{eq:covarianceMeff}
\begin{align}
\frac{\mu  s_1 s_4}{4}&=(\partial_{s_2}M_{\rm eff})\partial_{s_2}\partial_{s_4}M_{\rm eff}-(\partial_{s_4}M_{\rm eff})\partial_{s_2}^2M_{\rm eff},\label{eq:covarianceequationgeneral1}\\
0=&(\partial_{s_2}\mu)\partial_{s_4}M_{\rm eff}-(\partial_{s_2}M_{\rm eff})\partial_{s_4}\mu.\label{eq:covarianceequationgeneral2}
\end{align}
\end{subequations}

Equations \eqref{eq:HeffOnMeff} and \eqref{eq:covarianceMeff} provide the most general form of $\heff$ that depends solely on the basic scalars $s_a$ ($a=1,2,\cdots,5$), which are exactly the same scalars appearing in the Hamiltonian constraint of Einstein's GR. Such a dependence ensures that the Hamiltonian constraint contains at most second-order derivatives of the metric.

For classical GR, the mass function is $M_{\rm cl}=\frac{\sqrt{s_1}}{2}\left[1+(s_2)^2-(s_4)^2/4\right]$ which is a solution to Eq. \eqref{eq:covarianceMeff} with $\mu=1$. 
Besides, some other special solutions to the covariance equation were proposed in \cite{Zhang:2024khj,Zhang:2024ney}, leading to effective Hamiltonian constraints that depend on a quantum parameter. Models based on these special solutions have been extensively studied (see, e.g., \cite{Liu:2024wal,Konoplya:2024lch,Lin:2024beb,Liu:2024soc,
%Belfaqih:2024dzn,
Du:2024ujg,Konoplya:2025hgp,Ongole:2025lti,Motaharfar:2025ihv,Ai:2025myf,DelAguila:2025pgy}). 

Building on the above framework, two key open questions remain: how the dynamics behaves when $M_{\rm eff}$ and $\mathcal{R}$ are taken in their most general forms, and what the complete solution space of the covariance equation \eqref{eq:covarianceMeff} is. 
In this Letter, we systematically address these two problems and finally provide a comprehensive resolution of the inverse problem of Birkhoff's theorem.

\mysection{Birkhoff's theorem and dynamics for general $\heff$}
The constraint equations $\heff=0=H_x$ yield $M_{\rm eff}=M(t,x)$, where $M(t,x)$ satisfies
\begin{equation}\label{eq:heffvanishfinally}
\partial_x M+\mathcal R(E^1,M)\partial_x E^1=0.
\end{equation}
Moreover, considering the time evolution generated by $\heff[N]+H_x[N^x]$ for arbitrary lapse function $N$ and shift vector $N^x$, we can obtain
$\partial_t M_{\rm eff}+\mathcal R(E^1,M)\,\partial_t E^1=0$. This implies that the integration constant in the solution of Eq. \eqref{eq:heffvanishfinally} is independent of $t$. Namely, the integration constant is a Dirac observable of the model (see Appendix~\ref{app:A} for details).

As shown in \cite{Zhang:2024khj,Zhang:2024ney}, $\heff$ and $H_x$ form a closed algebra under the Poisson bracket, rendering the system first-class. Obtaining dynamics therefore requires gauge fixing. Consider an arbitrary gauge‐fixing that specifies the fields $E^I(t,x)$ for $I=1,2$. The lapse function $N$ and shift vector $N^x$ are then fixed by the evolution equations for $E^I$, i.e., $\dot E^I=\{E^I,\heff[N]+H_x[N^x]\}$, yielding (see Appendix \ref{app:A} for details)
\begin{equation}\label{eq:Nfinal}
\begin{aligned}
N^x=&\frac{\partial_t E^1}{\partial_xE^1}-\frac{2 N\partial_{s_2}M_{\rm eff}}{\partial_xE^1},\\
N=&C\frac{\partial_xE^1}{E^2}\exp(\int (\partial_{M_{\rm eff}}\mathcal R)\partial_xE^1\dd x ),
\end{aligned}
\end{equation}
where $C$ satisfies $\partial_x C=0$ on any open coordinate patch over which $\partial_{s_2}M_{\rm eff}=0$ is enforced by the gauge, while on patches with $\partial_{s_2}M_{\rm eff}\neq 0$ it obeys 
\begin{equation}\label{eq:dxC}
\begin{aligned}
\partial_xC=&\exp(-\int (\partial_{M_{\rm eff}}\mathcal R)\partial_xE^1\dd x )\times\\
&\frac{\partial_x\left(E^2\partial_tE^1\right)-\frac{E^2(\partial_tE^1)\partial_x^2E^1}{ \partial_xE^1}-(\partial_tE^2)\partial_xE^1}{2(\partial_{s_2}M_{\rm eff})\partial_xE^1}.
\end{aligned}
\end{equation}
In these expressions, $\partial_{M_{\rm eff}} \mathcal{R}$  denotes the partial derivative of $\mathcal{R}(s_1, M_{\rm eff})$ with respect to $M_{\rm eff}$, evaluated at $s_1 = E^1(t,x)$ and $M_{\rm eff} = M(t,x)$. The term $\partial_{s_2} M_{\rm eff}$ is understood analogously. 

 According to Eq. \eqref{eq:covarianceequationgeneral2}, $\mu$ is a function of $s_1$ and $M_{\rm eff}$. Hence, the dependence of $\mu$ on  $(t,x)$ can be obtained by substituting $s_1\equiv E^1(t,x)$ and $M_{\rm eff}=M(t,x)$. Applying this result, together with Eq. \eqref{eq:Nfinal} and the expressions of $E^I(t,x)$ specified for the gauge fixing, we can obtain the  $(t,x)$-dependence of the metric given in Eq. \eqref{eq:metric}. 

The preceding analysis shows that different gauge choices fix different lapse functions $N$ and shift vectors $N^x$. Because the theory is generally covariant, the metrics corresponding to these choices are related by diffeomorphism transformations.  Hence the spacetime geometry is uniquely fixed, with the only remaining freedom being the integration constant---the Dirac observable appearing in Eq. \eqref{eq:heffvanishfinally}.
This demonstrates the analogue of Birkhoff's theorem for the models associated with $\heff$. Since $\heff$ is derived by requiring covariance, the above results establish a fundamental link between covariance and Birkhoff's theorem.

\mysection{General solution to the covariance equation}To solve Eq. \eqref{eq:covarianceequationgeneral1}, let us introduce $\tilde s_4=(s_4)^2$, and $\F(s_1,M_{\rm eff})$ satisfying $\partial_{M_{\rm eff}}\F=8/(\mu  s_1)$. Then, Eq.~\eqref{eq:covarianceequationgeneral1} can be simplified to
\begin{equation}\label{eq:eq:covarianceequationgeneral1pp}
1=(\partial_{s_2}\F)\partial_{\tilde s_4}\partial_{s_2}M_{\rm eff}-(\partial_{\tilde s_4}\F)\partial_{s_2}^2M_{\rm eff},
\end{equation}
which leads to (see Appendix \ref{app:B})
\begin{equation}\label{eq:solMeff}
(\partial_{s_2}M_{\rm eff})^2=\frac{\mu s_1}{4}\left[(s_4)^2+\mathcal Z\right],
\end{equation}
for an arbitrary function $\mathcal Z$ of $s_1$ and $M_{\rm eff}$. The equivalence between Eq.~\eqref{eq:covarianceequationgeneral1} and Eq.~\eqref{eq:solMeff} follows directly from substituting the latter into the former.  Since Eq. \eqref{eq:solMeff} is a first-order differential equation, it admits an integration constant $\Xi(s_1,s_4)$.
Consequently, all the generally covariant theories considered here are characterized by four functions, namely $\mathcal Z$, $\mu$, $\mathcal R$, and $\Xi$. As demonstrated above, each such theory admits a unique class of static, spherically symmetric vacuum solutions characterized by the Dirac observable encoded in $M_{\rm eff}$. The explicit form of the metric is obtained by inserting Eq. \eqref{eq:solMeff} into Eq. \eqref{eq:Nfinal} and following the procedure described thereafter. On any open coordinate patch where the gauge choice $E^1=x^2$ and $\partial_{s_2}M_{\rm eff}=0$ is admissible, the line element reads
\begin{equation}\label{eq:finalmetric}
\dd s^2=\frac{1}{4}N_o^2\mathcal Z\dd t^2-\frac{4}{\mu\mathcal Z}\dd x^2+x^2\dd\Omega^2,
\end{equation}
where $N_o$ is 
\begin{equation}\label{eq:Nfinalo}
N_o=\exp(2\int x(\partial_{M_{\rm eff}}\mathcal R)\dd x ).
\end{equation}
Here we used $(E^2)^2=-4x^2/\mathcal Z$ obtained by evaluating Eq. \eqref{eq:solMeff} in the gauge.  In addition,  since $E^1$ in the gauge is time independent, the line element \eqref{eq:finalmetric} is static.

\mysection{reconstructing covariant dynamics from metric}Let us consider any static, spherically symmetric metric, such as the Schwarzschild one, the Hayward one, or those inspired by candidate QG theories. We can identify the functions $\mu$, $\mathcal{Z}$, and $\mathcal{R}$ by comparing the metric with Eq.~\eqref{eq:finalmetric}. Then, substituting $\mu$ and $\mathcal{Z}$ into Eq. \eqref{eq:solMeff} determines $M_{\rm eff}$ as a function of $s_1$, $s_2$, and $s_4$. This $M_{\rm eff}$, together with the expression of $\mathcal{R}$, can subsequently be inserted into Eq.~\eqref{eq:HeffOnMeff} to obtain $\heff$. In this way, the generally covariant effective dynamics for the given metric is reconstructed, as detailed in what follows.

Consider a one-parameter family of static, spherically symmetric metrics
\begin{equation}\label{eq:generalSch}
\dd s^2 = -F(x;m) \dd t^2 + H(x;m)^{-1} \dd x^2 + x^2 \dd\Omega^2,
\end{equation}
where $F$ and $H$ are arbitrary functions of $x$, each depending on a mass parameter $m$. Comparing Eq.~\eqref{eq:generalSch} with Eq.~\eqref{eq:finalmetric}, we find 
\begin{equation}\label{eq:FHNZmu}
F=-\frac{1}{4}N_o^2\mathcal Z,\quad H=-\frac{1}{4}\mu\mathcal Z.
\end{equation}
Since Eq.~\eqref{eq:FHNZmu} provides two equations for three unknowns, $\mu$, $\mathcal{Z}$, and $\mathcal{R}$, one function must be specified as input.
We take $\mathcal{R}$ as input, after which $\mu$ and $\mathcal{Z}$ are uniquely determined as shown below.

As $\mathcal{R}$ is specified, Eq. \eqref{eq:heffvanishfinally} introduces an integration constant.
Identifying this integration constant with the mass parameter $m$, and using the previously established relation $M_{\rm eff} = M(x)$, we can solve for $m$ as a function of $M_{\rm eff}$ and $x$,
\begin{equation}\label{eq:defineA}
m=A(M_{\rm eff},x)
\end{equation}
which defines a function $A$. Here $M$ depends only on $x$  because the metric is static in the present gauge. 
Inserting the result Eq. \eqref{eq:defineA} into Eq. \eqref{eq:FHNZmu} and applying $s_1=x^2$, we finally have
\begin{equation}\label{eq:muzfina}
\begin{aligned}
\mu&=\frac{N_o(\sqrt{s_1})^2H(\sqrt{s_1};A(M_{\rm eff},\sqrt{s_1}))}{F(\sqrt{s_1};A(M_{\rm eff},\sqrt{s_1}))},\\
\mu\mathcal Z&=-4H(\sqrt{s_1};A(M_{\rm eff},\sqrt{s_1})).
\end{aligned}
\end{equation} 
Here, $N_o$ as a function of $\sqrt{s_1}=x$ is given by Eq.~\eqref{eq:Nfinalo} with $\mathcal R$ taken as the previously specified input function. Combining Eqs.~\eqref{eq:muzfina} and \eqref{eq:solMeff}, we finally have
\begin{equation}\label{eq:determineMas124}
\begin{aligned}
&\pm \frac{\sqrt{s_1}}{2}s_2+\Xi(s_1,s_4)\\
=&\int^{M_{\rm eff}}\dd \xi \Bigg\{\frac{(s_4)^2N_o(\sqrt{s_1})^2H(\sqrt{s_1};A(\xi,\sqrt{s_1}))}{F(\sqrt{s_1};A(\xi,\sqrt{s_1}))}\\
&-4H(\sqrt{s_1};A(\xi,\sqrt{s_1}))\Bigg\}^{-1/2}.
\end{aligned}
\end{equation}
From Eq. ~\eqref{eq:determineMas124}, we can solve for $M_{\rm eff}$ as a function of $s_1$, $s_2$ and $s_4$ explicitly. Then, substituting it into Eq.~\eqref{eq:HeffOnMeff} yields the expression of $\heff$, which governs the dynamics of the model. 
 
In what follows, we show explicit forms of $M_{\rm eff}$ for several concrete examples.  As the first example, we consider the classical Schwarzschild spacetime with 
\begin{equation}\label{eq:FSCH}
H=F=1-\frac{2m}{x}.
\end{equation}
Choosing  $\mathcal R=0$ and substituting Eq. \eqref{eq:FSCH} into Eq. \eqref{eq:determineMas124}, we finally have
\begin{equation*}
M_{\rm eff}=\frac{\sqrt{s_1}}{2}\left[1+(s_2)^2-\frac{(s_4)^2}{4}\right]+2 \frac{\Xi ^2}{\sqrt{s_1}}\pm 2 s_2 \Xi. 
\end{equation*}
To reproduce Einstein's GR, we can choose $\Xi = 0$, while other choices yield inequivalent covariant theories that nevertheless admit the Schwarzschild metric as their unique vacuum solution.

As the second example, we consider the loop quantum black hole spacetime with 
\begin{equation}\label{eq:HFloop}
H=F=1-\frac{2m}{x}\left(1-\frac{2\zeta^2m}{x^3}\right).
\end{equation}
Note that this metric was previously derived in, e.g., \cite{Kelly:2020uwj,Lewandowski:2022zce}, using different approaches, but always in the absence of a covariant dynamical framework.
Choosing $\mathcal R=0$ and plugging Eq. \eqref{eq:HFloop} into Eq. \eqref{eq:determineMas124}, we finally have
\begin{equation*}
\begin{aligned}
M_{\rm eff}=&\frac{\sqrt{s_1}^3}{2 \zeta ^2} \sin^2 \left(\frac{\zeta  s_2}{\sqrt{s_1}}\pm \frac{2\zeta  \Xi }{s_1}\right)\\
&\mp \frac{s_1 \sqrt{(s_4)^2-4} \sin \left(\frac{2 \zeta  s_2}{\sqrt{s_1}}\pm \frac{4 \zeta  \Xi}{s_1}\right)}{4 \zeta }.
\end{aligned}
\end{equation*}
To ensure the correct classical limit, i.e., that $M_{\rm eff}$ returns to the expression in Einstein's  GR as $\zeta$ approaches $0$, we should choose $\Xi= \sqrt{s_1(s_4)^2-4s_1}/4.$ 

For the Hayward metric \cite{Hayward:2005gi}, we have
\begin{equation}\label{eq:fhhay}
F(x;m)=H(x;m)\equiv F_H(x;m)=1-\frac{2m x^2}{x^3+2\zeta^2m}. 
\end{equation}
With choosing $\mathcal R=0$, we finally have 
\begin{equation*}
\begin{aligned}
&-\frac{\left(2 \zeta ^2 M_{\rm eff}+\sqrt{s_1}^{3}\right) \sqrt{(s_4)^2-4F_H(\sqrt{s_1};M_{\rm eff})}}{8 s_1+2 \zeta ^2 \left((s_4)^2-4\right)}\\
=&\frac{2 \sqrt{s_1}^5 \text{arctanh}\left(\frac{\zeta  \sqrt{(s_4)^2-4F_H(\sqrt{s_1};M_{\rm eff})}}{\sqrt{4 s_1+\zeta ^2 \left((s_4)^2-4\right)}}\right)}{\zeta  \sqrt{4s_1+\zeta ^2 \left((s_4)^2-4\right)}^{3}}\mp \frac{1}{2}  \sqrt{s_1} s_2,
\end{aligned}
\end{equation*}
which defines $M_{\rm eff}$ as a function of $s_1$, $s_2$ and $s_4$ implicitly. In the Hayward case, we have chosen $\Xi=0$ to ensure the classical limit. 

\mysection{Conclusion and discussion}The previous work  \cite{Zhang:2024khj,Zhang:2024ney} analyzed how general covariance restricts theories. 
Derived from first principles, the analysis is fully general and not tied to any specific model or quantization scheme. It established the  most general form of the effective Hamiltonian constraint under the assumptions of 1) the diffeomorphism constraint retaining its classical form; 2)  the given structure of the constraint algebra;  and  3) the Hamiltonian constraint depending on the same set of basic scalars $s_a$ as in Einstein's GR. This framework encompasses all generally covariant gravity theories consistent with these assumptions.
Within this framework, the present work establishes the following foundational results that go beyond previous analyses.

First, the considered generally covariant theories are characterized by four functions, $\mathcal Z$, $\mu$, $\mathcal R$, and $\Xi$, and each admits a unique family of static, spherically symmetric vacuum solutions, characterized by the Dirac observable associated with the black hole mass. This demonstrates that Birkhoff's theorem holds universally within this class of theories.

Second, we develop a systematic and model-independent procedure to reconstruct covariant dynamics consistent with Birkhoff's theorem from a one-parameter family of static, spherically symmetric geometries, where the resulting generally covariant theory is uniquely determined once the functions $\mathcal R$ and $\Xi$ are specified. This reconstruction framework establishes the covariant origin of any static, spherically symmetric spacetime, including the observationally inferred, quantum-gravity-inspired, and regular black holes, and enables observation-based tests of gravitational dynamics. Although current observations primarily involve rotating black holes, each admits a well-defined spherically symmetric limit; for example the Kerr geometry reduces to the Schwarzschild as the spin vanishes. This correspondence allows gravitational-wave and shadow measurements to probe and constrain the corresponding families of covariant theories.

Finally, the success of this approach motivates its extension to less symmetric settings, such as axisymmetric configurations and ultimately the full theory. Importantly, because the present Hamiltonian constraint represents the most general form consistent with the stated assumptions, any full covariant theory satisfying the counterparts of these assumptions will reduce to this framework in the spherically symmetric sector.

\begin{acknowledgements}
This work was supported in part by the National Key Research and Development Program of China Grant No. 2021YFC2203001 and in part by the NSFC (Grants No. 12275022, No.~12475046, and No. 12505055) and  ``the Fundamental Research Funds for the Central Universities''.
\end{acknowledgements}

\appendix
\begin{center}
\textbf{Appendices}
\end{center}
%%%%For Editorial 
%\section{Solving the dynamics}\label{app:A}
\setcounter{equation}{0}
\renewcommand{\theequation}{A\arabic{equation}}
\mylabel{}\label{app:A}\myappendix{Solving the dynamics}Equation \eqref{eq:heffvanishfinally} can be verified easily using the identity \cite{Zhang:2024ney}
\begin{equation*}
\heff-2\frac{\partial_{s_2}M_{\rm eff}}{\partial_xE^1} H_x=-\frac{2E^2}{\partial_xE^1}\left(\mathcal R\partial_x E^1+\partial_xM_{\rm eff}\right).
\end{equation*}
Moreover, substituting this identity into the evolution equation $\partial_t M_{\rm eff}=\{M_{\rm eff},\heff[N]+H_x[N^x]\}$ and imposing $H_x=0$, we obtain
\begin{equation}\label{eq:dtM1}
\partial_tM_{\rm eff}+2N\mathcal R\partial_{s_2}M_{\rm eff}=N^x\partial_xM_{\rm eff}.
\end{equation}
Additionally, with $H_x=0$ applied, the evolution equations for $E^I$ read
\begin{equation}\label{eq:dotE1general}
\begin{aligned}
\dot E^1&=2N\partial_{s_2}M_{\rm eff}+N^x\partial_xE^1,\\
\dot E^2&=\partial_x(N^xE^2)+N\frac{2E^2}{\partial_xE^1}\partial_x(\partial_{s_2}M_{\rm eff})\\
&+2NE^2(\partial_{M_{\rm eff}}\mathcal R)\partial_{s_2}M_{\rm eff}.
\end{aligned}
\end{equation}
The first line gives a relation between $N$ and $N^x$. Inserting this relation into Eq. \eqref{eq:dtM1} and using Eq. \eqref{eq:heffvanishfinally}, we find $\partial_t M_{\rm eff}+\mathcal R\partial_t E^1=0$.

Let us impose a gauge fixing that specifies $E^I(t,x)$ ($I=1,2$). On any coordinate patch where $\partial_{s_2}M_{\rm eff}\neq 0$, the evolution equation \eqref{eq:dotE1general} determines $N$ and $N^x$, yielding Eqs. \eqref{eq:Nfinal} and \eqref{eq:dxC}. On patches where $\partial_{s_2}M_{\rm eff}= 0$, Eq. \eqref{eq:dotE1general} no longer fixes $N$. In this case, the evolution equation of $\partial_{s_2}M_{\rm eff}$ is needed. A direct calculation shows 
\begin{equation*}
\begin{aligned}
&\partial_xN-N\left((\partial_xE^1 )\partial_{M_{\rm eff}}\mathcal R+\partial_x\ln(\frac{\partial_xE^1}{E^2})\right)\\
=&\frac{2(E^2)^2}{\mu E^1\partial_xE^1}\left(\partial_t\left(\partial_{s_2}M_{\rm eff}\right)-\frac{N\partial_x\left[\left(\partial_{s_2}M_{\rm eff}\right)^2\right]}{\partial_xE^1}\right)\\
=&0,
\end{aligned}
\end{equation*}
where the first equality is obtained by substituting Eqs. \eqref{eq:covarianceequationgeneral1} and \eqref{eq:heffvanishfinally} and by the vanishing of $\{\partial_{s_2}M_{\rm eff},H_x[N^x]\}=N^x\partial_x(\partial_{s_2}M_{\rm eff})$ on patches with $\partial_{s_2}M_{\rm eff}=0$. This result leads to Eq. \eqref{eq:Nfinal} with $\partial_xC=0$. 

%%%For Editorial 
%\section{Solving the covariance equation}\label{app:B}
\setcounter{equation}{0}
\renewcommand{\theequation}{B\arabic{equation}}
\mylabel{}\label{app:B}\myappendix{Solving the covariance equation}Let us start from Eq.~\eqref{eq:eq:covarianceequationgeneral1pp}. By defining $Q=\partial_{s_2}M_{\rm eff}$, we can rewrite it as
\begin{equation*}
\begin{aligned}
\det
\begin{pmatrix}
\partial_{s_2}\F&\partial_{\tilde s_4}\F\\
\partial_{s_2}Q&\partial_{\tilde s_4}Q
\end{pmatrix}=1.
\end{aligned}
\end{equation*}
This result, combined with the properties of the Jacobian matrix, leads to
\begin{equation*}
\begin{aligned}
\begin{pmatrix}
\partial_{\F}s_2&\partial_Q{s_2}\\
\partial_\F{\tilde s_4}&\partial_Q{\tilde s_4}
\end{pmatrix}
=\begin{pmatrix}
\partial_{\tilde s_4}Q&-\partial_{\tilde s_4}\F\\
-\partial_{s_2}Q&\partial_{s_2}\F
\end{pmatrix}.
\end{aligned}
\end{equation*}
Then, we have
\begin{equation*}
\frac{\partial \tilde s_4}{\partial Q}=\partial_{s_2}\F=Q\partial_{M_{\rm eff}}\F,
\end{equation*}
leading to Eq.~\eqref{eq:solMeff} ultimately. 

%\mycolor{
%\setcounter{equation}{0}
%\renewcommand{\theequation}{C\arabic{equation}}
%\mylabel{}\label{app:C}\myappendix{The metric in PG gauge}
%The explicit expression of the metric under the PG coordinates reads
%\begin{equation*}
%\begin{aligned}
%\dd s^2=&-N^2\dd t^2+\frac{1}{\mu}\left(\dd x\mp N\sqrt{\mu\left(1+\frac{1}{4}\mathcal Z\right)} \dd t\right)^2+x^2\dd\Omega^2.
%\end{aligned}
%\end{equation*}
%The coordinate transformation between the Schwarzschild time $t_s$ and the PG time $t$ reads
%\begin{equation*}
%\dd t=\dd t_s\pm \frac{4 \sqrt{\mu \left[\frac{1}{4}\mathcal Z+1\right]}}{\mu \mathcal Z N}\dd x.
%\end{equation*}
%}

\bibliography{reference}
\end{document}